\documentclass[10pt,conference]{IEEEtran}

\useRomanappendicesfalse

\usepackage[T1]{fontenc}
\usepackage[ansinew]{inputenc}
\usepackage{amssymb,amsmath,amsfonts,amscd, mathrsfs}

\def\A{ \mathcal{A} }
\def\D{ \mathcal{D} }
\def\AD{ \mathcal{A}(\mathcal{D}) }
\def\Hinf{ \mathcal{H}^{\infty}(\mathcal{D}) }

\newtheorem{theorem}{\bfseries Theorem}
\newtheorem{lemma}[theorem]{\bfseries Lemma}
\newtheorem{corollary}[theorem]{\bfseries Corollary}

\begin{document}

\title{Spectral Factorization, Whitening- and Estimation Filter -- Stability, Smoothness Properties and FIR Approximation Behavior}

\author{\authorblockN{Holger Boche and Volker Pohl}
        \authorblockA{Technical University Berlin, Heinrich Hertz Chair for Mobile Communications\\                     
                      Einsteinufer 25, 10587 Berlin, Germany,\quad
                      Email: \{holger.boche, volker.pohl\}@mk.tu-berlin.de}
}

\maketitle

\begin{abstract}
A Wiener filter can be interpreted as a cascade of a whitening- and an estimation filter. This paper gives a detailed investigates of the properties of these two filters. Then the practical consequences for the overall Wiener filter are ascertained. 
It is shown that if the given spectral densities are smooth (Hoelder continuous) functions, the resulting Wiener filter will always be stable and can be approximated arbitrarily well by a finite impulse response (FIR) filter. Moreover, the smoothness of the spectral densities characterizes how fast the FIR filter approximates the desired filter characteristic. If on the other hand the spectral densities are continuous but not smooth enough, the resulting Wiener filter may not be stable.
\end{abstract}

\section{Introduction}

The solution of the the optimal smoothing and prediction problem was published by Wiener, Hopf and Kolmogoroff \cite{Wiener_Hopf_1931,Kolmogorov_1941} in the 1940's. The well known formal solution of this problem (the causal Wiener filter) is determined from two given spectral densities with certain properties. The question is, how does these properties pass over to the Wiener filter, for instance: to the smoothness of the solution \cite{Wang_90}, as a mathematical attribute; to the stability of the filter, as a practical characteristic; or to the possibility to approximate the Wiener solution by an FIR filter, as an important practical and mathematical property. The answer of these questions is non-trivial because the operations, necessary to calculate the \emph{causal} Wiener filter, are non-linear.

In this paper the problem is completely solved for the most important smoothness class of spectral densities, namely for Hoelder continuous functions. Moreover, it is shown that the results hold not for continuous spectra, in general. In this paper, all investigations are done in the energy norm, but a similar treatment in the bounded-input bounded-output norm, which yields completely different results, was done in \cite{Boche_Pohl_CWIT05}.

\section{Notations \& Problem Statement}
\label{sec:Problem}

\subsection{Some system theoretical aspects}

Consider a time invariant and time-discrete linear system $\mathcal{S}$.
If $\{h_{k}\}$ is its impulse response then the input-output relation $y = \mathcal{S} x$ in the time domain is
$y_{n} = \sum^{\infty}_{k=-\infty} h_{k}x_{n-k}$
in which $x_{n}$ is an input- and $y_{n}$ is the corresponding output sequence. 
If $h_{k}=0$ for all $k<0$, the system is called \emph{causal}.
The so called \emph{$\mathscr{D}$-transform} of an arbitrary sequence $\{x_{k}\}$ and the corresponding inverse $\mathscr{D}$-transform is defined by
\begin{equation}
\label{equ:D_Transform}
	x(z) = \!\!\! \sum^{\infty}_{k=-\infty} \!\! x_{k} z^{k}
	\!\!\quad \textrm{and} \!\!\quad
	x_{k} = \frac{1}{2\pi\imath}\oint_{\left|z\right|=1}\!\!\!\!x(z) z^{-(k+1)}dz\ .
\end{equation}
Note that the $\mathscr{D}$-transform is the usual $\mathscr{Z}$-transform where $z$ is replaced by $z^{-1}$. The $\mathscr{D}$-transform of an impulse response $\{h_{k}\}$ is called the \emph{transfer function} $h$ of the linear system and its \emph{frequency response} $h(e^{-\imath\omega})$ with $\omega\in [-\pi,\pi)$ is obtained by setting $z=e^{-\imath\omega}$ in its transfer function.
From the definition of the $\mathscr{D}$-transform \eqref{equ:D_Transform} and from the Cauchy integral theorem, it is clear that $\{h_{k}\}$ is causal if the corresponding transfer function $h(z)$ is an analytic function inside the unit disk $\mathcal{D}:=\left\{z \in \mathbb{C} : \left|z\right|<1\right\}$.

In the $\mathscr{D}$-domain the input-output relation becomes $y(e^{-\imath\omega}) = h(e^{-\imath\omega})x(e^{-\imath\omega})$ for $\omega \in \left[-\pi,\pi\right)$.
The average energy of a sequence $x_{n}$ is given by the square of its $l^{2}$- resp. $L^{2}$-norm $\left\|x\right\|_{2}$.
The \emph{stability norm} of a linear system $\mathcal{S}$ is defined by
\begin{eqnarray*}
	\left\|\mathcal{S}\right\|_{E} := \sup_{\left\|x\right\|_{2}\leq 1}\left\|\mathcal{S}x\right\|_{2}
\end{eqnarray*}
and it is well known that $\left\|\mathcal{S}\right\|_{E} = \left\|h\right\|_{\infty}$ where $\left\|h\right\|_{\infty} = \sup_{\omega} \left|h(e^{-\imath\omega})\right|$ is the usual supremum norm of the frequency response of $\mathcal{S}$. For a causal $h$ this norm can also be written as $\left\|h\right\|_{\infty}=\sup_{\left|z\right| < 1}\left|h(z)\right|$.
Any linear system $\mathcal{S}$ is called (energy) \emph{stable} if it has a finite energy norm $\left\|\mathcal{S}\right\|_{E}<\infty$, i.e. if any input $x$ with $\left\|x\right\|_{2} < \infty$ yields an output $y$ with $\left\|y\right\|_{2} < \infty$.

Now, all causal and stable transfer functions are collected in the \emph{Hardy space} $\Hinf$. This is the set of all complex functions which are analytic and bounded (with respect to $\left\|\cdot\right\|_{\infty}$) in $\D$. Together with the point wise multiplication, $\Hinf$ is a \emph{Banach algebra}.

Consider now a transfer function $h_{N}$ of the form $h_{N}(z) = \sum^{N}_{k=0} h_{k}z^{k}$. Practically, the corresponding $\mathcal{S}_{N}$ can be realized by an FIR filter of degree $N$. We say that a linear system $\mathcal{S}$ \emph{can be approximated uniformly by an FIR filter} if there exists a $\mathcal{S}_{N}$ with transfer function $h_{N}$ such that the maximal difference between the output $y_{N} = \mathcal{S}_{N}x$ of the FIR filter and the output $y=\mathcal{S}x$ of the linear system becomes arbitrarily small as $N$ tends to infinity, i.e. if
\begin{equation*}
	\left\|\mathcal{S} - \mathcal{S}_{N}\right\|_{E} = 
	\sup_{\left\|x\right\|_{L^{2}}\leq 1} \left\|\left(h-h_{N}\right)x\right\|_{\infty} \rightarrow 0
	\quad \textrm{as} \quad
	N \rightarrow \infty.
\end{equation*}
It is known, that any $h \in \Hinf$ which is continuous in the closure of the unit disk $\overline{\D}$ can be approximated uniformly by an FIR filter. The sub-algebra of $\Hinf$ which contains these functions is the \emph{disk algeba} $\AD\subset\Hinf$. This is the set of all functions analytical in $\D$ and continuous in $\overline{\mathcal{D}}$. $\AD$ comprises all stable and causal transfer functions which can be approximated by FIR filters.

Since we consider discrete-time systems, the frequency responses are defined on the unit circle, i.e. they are defined on $[-\pi,\pi)$.  Therefore, we denote by $\mathcal{C}[-\pi,\pi)$ the set of all complex valued bounded and continuous functions on the interval $[-\pi,\pi)$ with $f(-\pi) = \lim_{\omega\rightarrow\pi}f(\omega)$
\footnote{Throughout this paper, this condition is always implicitly assumed, if a function space over the interval $[-\pi,\pi)$ is considered}.
To control the smoothness of our functions, we use the so called \emph{Hoelder spaces} (see e.g. \cite{Zygmund}): Let $\Omega$ be a domain in the complex plain. The set of all functions $f$ for which
\begin{equation*}
	\left\|f\right\|_{\alpha} := \left|f(0)\right| + \sup_{z_{1} \neq z_{2}} \frac{ \left|f(z_{1}) - f(z_{2})\right| }
	                             { \left|z_{1} - z_{2}\right|^{\alpha} } < \infty,
	                             \quad \forall z_{1},z_{2} \in \Omega
\end{equation*}
is denoted by $\mathcal{C}_{\alpha}(\Omega)$. This is the set of all $f$ which satisfy a Hoelder condition of order $\alpha$.
In particular, we need the spaces $\mathcal{C}_{\alpha}[-\pi,\pi)$ and $\A_{\alpha}(\D)$ with $0<\alpha< 1$. The latter is the subset of $\AD$ which elements satisfy a Hoelder condition of order $\alpha$. With the above norm they are Banach algebras.

The set of all real valued functions $f(\omega)$ defined on an interval $\Omega$ with $f(\omega)>0$ for all $\omega \in \Omega$ is denoted by $\mathcal{C}_{+}(\Omega)$.

\subsection{Wiener filtering}

The Wiener filtering problem can be formulated as follows \cite{Bode_Shannon_1950}: Given a wide sense stationary random process $\{y_{n}\}$ with power spectrum $\Phi$ as the input of a linear filter $H$. Let $\{x_{n}\}$ be the desired output of this filter and denote by $\Psi$ the Fourier transform of the cross-correlation function between the sequences $x_{n}$ and $y_{n}$.
We look for a causal $H$ such that the output of this filter
$\widehat{x}_{n} = \sum^{\infty}_{k=0} H_{k}y_{n-k}$
minimizes the mean square error (MSE) $\overline{\epsilon} = \mathcal{E}[\left|\widehat{x}_{k}-x_{k}\right|^{2}]$
to the desired sequence $x_{n}$, where $\mathcal{E}\left[\cdot\right]$ stands for expectation.

The formal solution to this filtering problem is well known, and the frequency response of the optimal Wiener filter is
\begin{equation}
\label{equ:WienerSolution}
	H(e^{-\imath\omega}) = \frac{1}{\Phi_{+}(e^{-\imath\omega})} \left[\frac{\Psi}{\Phi_{-}}\right]^{+}(e^{-\imath\omega})\ .
\end{equation}
The functions $\Phi_{+}$ and $\Phi_{-}$ are obtained by the \emph{spectral factorization} of the power spectrum $\Phi$ such that
\begin{equation}
\label{equ:Problem_Factorization}
	\Phi(\omega) = \Phi_{+}(e^{-\imath\omega})\Phi_{-}(e^{-\imath\omega})\ ,\quad \forall \omega\in\left[-\pi,\pi\right)
\end{equation}
and such that $\Phi_{+}$ is an analytic function in $\D$ without any zero inside $\D$ and such that $\Phi_{-}$ is an analytic function outside $\overline{\D}$ with $\Phi_{-}(z)\neq 0$ for all $\left|z\right|>1$. The symbol $[\cdot]^{+}$ in \eqref{equ:WienerSolution} is used to indicate the \emph{positive-time part} of the sequence whose Fourier-transform is inside the brackets. So let $f$ be a given transfer function with Fourier coefficients $f_{k}$,
then the positive-time part $\left[f\right]^{+}$ is given by $f^{+}(z) = \sum^{\infty}_{k=0}f_{k} z^{k}$.

Solution \eqref{equ:WienerSolution} is typical interpreted as a cascade of two filters 
\begin{equation}
\label{equ:Wiener_Cascade}
	H(e^{-\imath\omega}) = H_{W}(e^{-\imath\omega}) \cdot H_{E}(e^{-\imath\omega})
\end{equation}
with the two causal filters
\begin{equation}
\label{equ:ElementarFilter}
	H_{W}(e^{-\imath\omega}) = \frac{1}{\Phi_{+}(e^{-\imath\omega})}
	\!\!\quad \textrm{,} \!\!\quad
	H_{E}(e^{-\imath\omega}) = \left[\frac{\Psi}{\Phi_{-}}\right]^{+}(e^{-\imath\omega})
\end{equation}
The \emph{whitening filter} $H_{W}$ generates at its output a white noise process, whereas the succeeding \emph{estimation filter} $H_{E}$ produces the desired minimum-MSE solution $\widehat{x}$ from the whitened signal.

To determine the causal Wiener filter \eqref{equ:WienerSolution}, we have to solve two problems: First, the spectral factorization \eqref{equ:Problem_Factorization} of $\Phi$, and secondly the positive time part $\Gamma^{+}$ of the function $\Gamma:=\Psi/\Phi_{-}$ has to be calculated.
The formal solutions of both problems are well known in the engineering literature (see e.g. \cite{Hayes}). 
The spectral factorization \eqref{equ:Problem_Factorization} is given by
\begin{equation}
\label{equ:Factorization}
	\Phi_{+}(z)=\left[F_{+}(z)\right]^{1/2}
	\qquad \textrm{and}\qquad
	\Phi_{-}(z)=\overline{\Phi_{+}(1/z)}
\end{equation}
for all $z  \in  \D$ and in which $F_{+}$, given by
\begin{equation}
\label{equ:F+}
	F_{+}(z) = \exp\left(\frac{1}{2\pi}\int^{\pi}_{-\pi}\!\!\log \Phi(\tau)\ 
	\frac{e^{\imath\tau}+z}{e^{\imath\tau}-z}\ d\tau\right)
\end{equation}
is called \emph{outer function} of $\Phi$, and $\Phi_{\pm}$ are called the \emph{spectral factors} of $\Phi$.
The positive time part of $\Gamma$ is by definition
$\Gamma^{+}(z) = \sum^{\infty}_{k=0}\gamma_{k}z^{k}$. Inserting the Fourier coefficients $\gamma_{k}$ \eqref{equ:D_Transform} of $\Gamma$ gives the closed form solution for all $z \in \D$
\begin{equation}
\label{equ:TimeDecomposition_f+}
	\Gamma^{+}(z) = \frac{1}{2\pi} \int^{\pi}_{-\pi} \Gamma(\tau) \frac{e^{\imath\tau}}{e^{\imath\tau} - z}\ d\tau\ .
\end{equation}

It should be noted that $\Gamma^{+}$ exits if $\Gamma$ is absolute integrable.
The spectral factorization has a solution, if $\Phi$ is absolute integrable and satisfies the  Paley-Wiener condition.
Note the very similar structure of \eqref{equ:F+} and \eqref{equ:TimeDecomposition_f+}.
The solution \eqref{equ:TimeDecomposition_f+} is given by the Cauchy transform of $\Gamma$ with the kernel $\sum^{\infty}_{k=0}[ze^{-\imath\tau}]^{k} = e^{\imath\tau}/(e^{\imath\tau} - z)$. The solution \eqref{equ:F+} is given by the exponential function of an integral transformation of the logarithm of $\Phi$, wherein the kernel is $\left(1 + \sum^{\infty}_{k=1}[ze^{-\imath\tau}]^{k}\right)/2 = (e^{\imath\tau} + z)/(e^{\imath\tau} - z)$ and therefore very similar to the kernel of the Cauchy transform.
Nevertheless, even although both solutions have such a similar formal structure, the behavior of both solutions is very different, as shown in this paper.

Therewith, the solution of the Wiener filtering problem is formally completely solved. From a practical point of view, this solution has to satisfy some desired properties such as stability or the possibility to approximate it by an FIR filter. Subsequently, we study which conditions $\Phi$ and $\Psi$ have to satisfy such that the Wiener filter is stable and can be approximated by an FIR filter. To this end, the two filters $H_{W}$ and $H_{E}$ are considered separately at first.

\section{Properties of the whitening filter}
\label{sec:WhiteningFilter}

In this section, the properties of the pre-whitening filter $H_{W}$ are investigated. Its formal solution is given by
\begin{equation}
\label{equ:whiteningFilter}
	H_{W}(z) = \left[F_{+}(z)\right]^{-\frac{1}{2}}
\end{equation}
in which $F_{+}$ is defined by \eqref{equ:F+} in terms of the given $\Phi$.

First we consider the case that $\Phi$ satisfies a certain Hoelder condition of order $\alpha$. It turns out that the corresponding $H_{W}$ will be in $\AD$ and satisfies the same Hoelder condition as $\Phi$. So in this case, $H_{W}$ is a stable filter which can be approximated uniformly by an FIR filter.
Moreover, the order $\alpha$ of the Hoelder condition determines how fast the approximation error decreases as the degree of approximation polynomial increases (cf. Section~\ref{sec:FIR_Approximation}).
In the second part of this section, it is shown that there exist continuous functions $\Phi$ such that $H_{W}$ is not in $\AD$. Therefore it is not possible to approximate $H_{W}$ by an FIR filter in this case.

\subsection{Smooth spectral densities}

\begin{theorem}
\label{theo:SpectralFac_Smooth}
Let $\Phi \in \mathcal{C}_{\alpha}[-\pi,\pi)$ with $0<\alpha<1$ be a real valued function with $\Phi(\omega)\geq\delta>0$ for all $\omega\in [-\pi,\pi)$. Then for its outer function $F_{+}$ given by \eqref{equ:F+} holds: $F_{+} \in \A_{\alpha}(\D)$ and $\left|F_{+}(z)\right|\geq\delta>0$ for all $z\in \overline{\D}$.
\end{theorem}

The requirement that $\Phi$ should be strictly positive is a natural condition for our problem, because it ensures that $H_{W}$ always exists. It should be emphasized that for $\alpha=1$ the above theorem is not valid, in general.

Before this theorem is proofed, we state the consequences of this result concerning the spectral factors $\Phi_{\pm}$ and the whitening filter $H_{W}$. Since they are all functions of $F_{+}$ (see \eqref{equ:Factorization} and \eqref{equ:whiteningFilter}) and since $F_{+}$ is bounded and strictly positiv in $\overline{\D}$, it can easily be shown that $\Phi_{\pm}$ and $H_{W}$ belongs to same space $\A_{\alpha}(\D)$ as $F_{+}$.
Thus:

\begin{corollary}
\label{Corollary_SpectralFac_Smooth}
Let $\Phi \in \mathcal{C}_{\alpha}[-\pi,\pi)$ for $0<\alpha<1$ be a real valued function with $\Phi(\omega)\geq\delta>0$ for all $\omega\in [-\pi,\pi)$. Then $\Phi_{+}(z)$, $\Phi_{-}(1/z)$ and $H_{W}(z)$ belong to $\A_{\alpha}(\D)$ and it holds $\left|\Phi_{+}(z)\right| \geq \sqrt{\delta}$  and $\left|\Phi_{-}(1/z)\right| \geq \sqrt{\delta}$ for all $\left|z\right| \leq 1$.
\end{corollary}

This shows that if $\Phi$ is Hoelder continuous, the spectral factors $\Phi_{\pm}$ and the whitening filter $H_{W}$ are stable and can be approximated arbitrarily well by FIR filters. Moreover, they satisfy the same Hoelder condition as $\Phi$.

\begin{proof}(of Theorem~\ref{theo:SpectralFac_Smooth})
From the assumptions on $\Phi$
follows immediately that $\log \Phi \in \mathcal{C}_{\alpha}[-\pi,\pi)$. Consider then $F_{+}$ as defined in (\ref{equ:F+}) and write it at $z=re^{\imath\omega}$ as
\begin{equation}
\label{equ:F+_ausfuehrlich}
    F_{+}(re^{\imath\omega}) = \exp\left(\ u(re^{\imath\omega}) + \imath\cdot v(re^{\imath\omega})\ \right)
\end{equation}
with
\begin{eqnarray}
		\label{equ:u}
    u(re^{\imath\omega})\!\!\!\!\! &=& \!\!\!\!\frac{1}{2\pi}\int^{\pi}_{-\pi}\!\!\!\log \Phi(\tau) 
    \frac{1-r^{2}}{1-2r\cos(\omega-\tau)+r^{2}}\ d\tau \\[4pt]
    \label{equ:v}
    v(re^{\imath\omega})\!\!\!\!\! &=& \!\!\!\!\frac{1}{2\pi}\int^{\pi}_{-\pi}\!\!\!\log \Phi(\tau) 
    \frac{2r\sin(\omega-\tau)}{1-2r\cos(\omega-\tau)+r^{2}}\ d\tau
\end{eqnarray}
The function $u$ is the Poisson integral of $\log \Phi$ and therefore $u$ is harmonic in $\D$ and continuous in $\overline{\D}$ with $u(e^{\imath\omega}) = \log \Phi(\omega)$ for all $\omega \in [-\pi,\pi)$. Now, it can be shown that $u$ satisfies in the whole $\overline{\D}$ the same Hoelder condition as $\log \Phi$ and therefore $u \in \mathcal{C}_{\alpha}(\overline{\D})$.
The function $v$ is the conjugate Poisson integral of $\log \Phi$ or equivalent $v$ is the Poisson integral of the Hilbert transform of $\log \Phi$. Next, we use a theorem of Privalov-Zygmund (see e.g. \cite[chapter~3]{Zygmund}) which states that if a function $\Phi$ belongs to $\mathcal{C}_{\alpha}[-\pi,\pi), 0<\alpha<1$, so does its Hilbert transform as well. For our case follows that the Hilbert transform of $\log \Phi$  belongs to $\mathcal{C}_{\alpha}[-\pi,\pi)$ and therefore with the same arguments as for $u$: $v \in \mathcal{C}_{\alpha}(\overline{\D})$.
Since $u$ and $v$ are harmonic functions in $\D$ and Hilbert transforms of each other, the complex function $g_{+}:=u+\imath\cdot v$ is analytic in $\D$ and consequently $g_{+}\in\A_{\alpha}(\D)$.
The outer function is obtained by $F_{+} = \exp( g_{+} )$. Of course, $F_{+}$ is also analytic in $\D$ and continuous in $\overline{\D}$ and it is easily verified that
$F_{+}$ satisfies also a Hoelder condition of order $\alpha$ such that indeed $F_{+} \in \A_{\alpha}(\D)$.\\
Finally, $\left|F_{+}\right| = \exp\left(u\right)$ with $u$ given by the Poisson integral \eqref{equ:u} of $\log \Phi$. Because $\Phi(\omega) \geq \delta$  for all $\omega\in [-\pi,\pi)$, it follows that $\log \Phi(\omega) \geq \log\delta$  and since $\Phi$ is continuous, we obtain that $\left|F_{+}(z)\right|\geq\delta$ for all $z \in \overline{\D}$.
\end{proof}

\subsection{Continuous spectral densities}

Now we show that the assumptions on the spectral density $\Phi$, made in the last sub-section, can not be weakened, i.e. if $\Phi$ is continuous but not Hoelder continuous, there exists such a $\Phi$ for which $F_{+}$ belongs not to $\AD$.

Since the given spectral density $\Phi\in \mathcal{C}_{+}[-\pi,\pi)$ is bounded, the outer function \eqref{equ:F+} is analytic and bounded in $\D$. This means that $F_{+}\in\Hinf$ represents a transfer function of a causal and stable filter. Moreover it is well known that every $F\in\Hinf$ has a radial limit $\lim_{r\rightarrow 1}F(re^{\imath\omega})$ almost everywhere (a.e.) in $[-\pi,\pi)$, and it was shown that the absolute value of $F_{+}$ is continuous in the whole $\overline{\D}$ with
\begin{eqnarray*}
  &  \lim_{r\rightarrow1}\left|F_{+}(re^{\imath\omega})\right| = \Phi(\omega) \quad , \quad
     \forall \omega \in \left[-\pi,\pi\right).
\end{eqnarray*}
Nevertheless, as the following theorem shows, the outer function itself is in general not continuous in $\overline{\D}$ and therefore not an element of $\AD$.

\begin{theorem}
\label{DiskAlgebra}
To any set $E$~$\subset$~$\left[-\pi,\pi\right)$ of Lebesgue measure zero there exists a function  $\Phi\in\mathcal{C}_{+}\left[-\pi,\pi\right)$ such that its corresponding outer function $F_{+}(re^{\imath\vartheta})$ has no limit for all $\vartheta\in E$ as $r\rightarrow1$.
\end{theorem}

In view of the spectral factorization and in respect to the properties of the whitening filter \eqref{equ:whiteningFilter} the following Corollary is deducted from this theorem.
\begin{corollary}
\label{Corollary2}
There exist spectral densities $\Phi\in \mathcal{C}_{+}\left[-\pi,\pi\right)$ such that there are no functions $\Phi_{+}(z)\in\AD$ and $\Phi_{-}(1/z) \in \AD$ which factorize $\Phi$ according to \eqref{equ:Problem_Factorization} and such that $H_{W}$ belongs not to $\AD$.
\end{corollary}

For the proof of Theorem~\ref{DiskAlgebra} the following lemma, which was already proofed in \cite{Boche_00_Cauchy}, will be necessary.
\begin{lemma}
\label{lem:Boche}
Let $E \subset \left[-\pi,\pi\right)$ be an arbitrary set of Lebesgue measure zero. There exits a function $h\in\mathcal{C}_{+}\left[-\pi,\pi\right)$ such that 
\begin{equation}
\label{equ:Lemma_Boche}
	\lim_{r\rightarrow1}\ \frac{1}{2\pi}\int^{\pi}_{-\pi}h(\omega)\ \frac{r\sin(\vartheta-\omega)}{1-2r\cos(\vartheta-\omega)+r^{2}}\ d\omega = -\infty
\end{equation}
for all $\vartheta \in E$.
\end{lemma}

\begin{proof}(Sketch of proof of Theorem~\ref{DiskAlgebra})
Let $E\in\left[-\pi,\pi\right)$ be an arbitrary set of Lebesgue measure zero and let $f(\omega):=\exp(h(\omega))$ for all $\omega\in [-\pi,\pi)$, where $h$ is a function corresponding to the set $E$ as in Lemma~\ref{lem:Boche}. Note that the so defined $f$ is a continuous and strictly positive function.
Consider now $F_{+}$ as defined in (\ref{equ:F+}) and write it at $z=re^{\imath\omega}$ as in \eqref{equ:F+_ausfuehrlich} with $u$ and $v$ as given by \eqref{equ:u} and \eqref{equ:v}.
Using Lemma~\ref{lem:Boche} for $v$, it can be shown that for all $\vartheta\in E$
\begin{eqnarray}
	\limsup_{r\rightarrow1}\ \Re\left\{F_{+}(re^{\imath\vartheta})\right\} &=& \Phi(\vartheta)\\
	\liminf_{r\rightarrow1}\ \Re\left\{F_{+}(re^{\imath\vartheta})\right\} &=& -\Phi(\vartheta)
\end{eqnarray}
And a completely similarly result holds for the imaginary part.
Since $\Phi(\omega)\neq 0$, this proofs that both $\Re\left\{F_{+}\right\}$ and $\Im\left\{F_{+}\right\}$ and therefore $F_{+}(\vartheta)$ itself diverges for all $\vartheta \in E$.
\end{proof}

Theorem~\ref{DiskAlgebra} states that there exists a $\Phi\in \mathcal{C}_{+}\left[-\pi,\pi\right)$ such that the corresponding $F_{+}$ belongs not to $\AD$. Since the spectral factors $\Phi_{\pm}$ and the whitening filter $H_{W}$ are determined by $F_{+}$ (see \eqref{equ:Factorization}, \eqref{equ:whiteningFilter}), Corollary~\ref{Corollary2} follows at once. This result has important practical consequences, because even if the spectral density $\Phi$ is continuous, it may not be possible to realize the desired filter characteristic of $H_{W}$ by an FIR filter with arbitrarily small error.

\section{Properties of the estimation filter}
\label{sec:EstimationFilter}

Now the properties of the estimation filter $H_{E}$, given by $\eqref{equ:ElementarFilter}$, are investigated. 
As we will show, there is a fundamental difference between the whitening filter $H_{W}$ and $H_{E}$. As it was shown, $H_{W}$ is always stable. However, the stability of $H_{E}$ depends strongly on the smoothness of the given spectral densities. If they satisfy no Hoelder condition, $H_{E}$ may become unstable.
First we consider again the case where $\Phi$ and $\Psi$ satisfy a certain Hoelder condition and show that $H_{E}$ is also a continuous function in $\overline{\D}$ and satisfies the same Hoelder condition as the given spectra. In the second part, spectra are considered which satisfy no Hoelder condition. It turns out that in this case, $H_{E}$ is even not bounded, in general.

Because of the limiting space, the proofs are omitted in this section. However, because of the similarity of the two operators \eqref{equ:F+} and \eqref{equ:TimeDecomposition_f+}, the techniques with that the results can be proofed is very similar to that used in Section~\ref{sec:WhiteningFilter}.

\subsection{Smooth spectral densities}

We assume that $\Phi \in \mathcal{C}_{\alpha}[-\pi,\pi)$ with $\Phi(\omega) > 0$ for all $\omega \in [-\pi,\pi)$. From this assumption follows (see Corollary~\ref{Corollary_SpectralFac_Smooth}) that the spectral factor $\Phi_{-}(1/z)$ belongs to $\A_{\alpha}(\D)$ and its absolute value is strictly positive in $\overline{\D}$. In particular $\Phi_{-}(e^{\imath\omega}) \in \mathcal{C}_{\alpha}[-\pi,\pi)$ with $\left| \Phi_{-}(e^{\imath\omega}) \right| > 0$ for all $\omega \in [-\pi,\pi)$. Therefore, also $1/\Phi_{-}(e^{\imath\omega})$ belongs to $\mathcal{C}_{\alpha}[-\pi,\pi)$.
Furthermore it is assumed that $\Psi \in \mathcal{C}_{\alpha}[-\pi,\pi)$. Since $\mathcal{C}_{\alpha}[-\pi,\pi)$ is an algebra, the quotient $\Gamma=\Psi / \Phi_{-}$ belongs also to $\mathcal{C}_{\alpha}[-\pi,\pi)$.
But to which function space belongs the positive time part of $\Gamma$? The answer is given by

\begin{theorem}
\label{theo:EstFilterSmooth}
Let $\Gamma \in \mathcal{C}_{\alpha}[-\pi,\pi)$ with $0<\alpha<1$. Then the positive time part $\Gamma^{+}$ belongs to $\A_{\alpha}(\D)$.
\end{theorem}

This theorem shows that as long as both spectra $\Phi$ and $\Psi$ satisfy a Hoelder condition, $H_{E}$ will be stable and can be approximated arbitrarily well by an FIR filter.

\subsection{Continuous spectral densities}

It was shown (Corollary~\ref{Corollary2}) that if $\Phi$ is only in $\mathcal{C}[-\pi,\pi)$, the corresponding spectral factor $\Phi_{-}$ may not be continuous in $\overline{\D}$. However, we assume that $\Gamma=\Psi / \Phi_{-}$ belongs to $\mathcal{C}[-\pi,\pi)$. For this case, we have 

\begin{theorem}
\label{theo:TimeDecomp_Continious}
To any set $E\subset[-\pi,\pi)$ of Lebesgue measure zero there exists a function $\Gamma \in \mathcal{C}[-\pi,\pi)$ such that $\left|\Gamma^{+}(re^{\imath\vartheta})\right|\rightarrow\infty$ as $r\rightarrow 1$ for all $\vartheta \in E$ and therefore $\left\|\Gamma^{+}\right\|_{\infty} = \infty$, i.e. $\Gamma^{+} \notin \Hinf$.
\end{theorem}

This theorem shows that if $\Phi$ and $\Psi$ satisfy no Hoelder condition, it can not be guaranteed that the estimation filter $H_{E}$ is stable in the energy sense.

\section{Properties of the Wiener Filter}
\label{sec:WienerFilter}

Now we consider the properties of the overall Wiener filter \eqref{equ:Wiener_Cascade}. They are easily obtained by combining the results of the two preceding sections. There, it was shown that if $\Phi$ and $\Psi$ are sufficiently smooth, i.e. satisfy a Hoelder condition of order $\alpha$, the whitening filter $H_{W}$ and the estimation filter $H_{E}$ will belong to $\A_{\alpha}(\D)$. Under point-wise multiplication, $\A_{\alpha}(\D)$ is an algebra. Therefore, also the Wiener filter $H=H_{W}\cdot H_{E}$ belongs to $\A_{\alpha}(\D)$. This important result is summarized in the following corollary:

\begin{corollary}
\label{cor:WienerFilter_smooth}
Let $\Phi,\Psi \in \mathcal{C}_{\alpha}\left[-\pi,\pi\right)$ then the transfer function $H$ of the corresponding Wiener filter belongs to $\A_{\alpha}(\D)$.
\end{corollary}

This corollary says that under the above assumptions on $\Phi$ and $\Psi$, the Wiener filter \eqref{equ:WienerSolution} is always stable and can be approximated arbitrary well by an FIR filter. Moreover, it makes a statement on the smoothness of $H$.
As it is shown in the next section, from the order $\alpha$ of the Hoelder condition the dependency of the approximation error from the FIR filter order can be determined.
If $\Phi$ and $\Psi$ are not Hoelder continuous, the Wiener filter my not be stable.
Also this result is summarized in a corollary:

\begin{corollary}
There exists continuous spectra $\Phi \in \mathcal{C}[-\pi,\pi)$ and $\Psi \in \mathcal{C}[-\pi,\pi)$ such that the Wiener filter $H$, given by \eqref{equ:WienerSolution}, is not stable, i.e. such that $H \notin \Hinf$.
\end{corollary}

Of course, because of Corollary~\ref{cor:WienerFilter_smooth}, at least one of $\Phi$ and $\Psi$ satisfies no Hoelder condition in this case.

Finally, it should be mentioned that all the results, derived here for continuous spectral densities which satisfy a Hoelder condition of order $\alpha$ (i.e. functions of $\mathcal{C}_{\alpha}$) can easily be extended to spectral densities which are $k$-times continuous differentiable, and whose $k$-th derivate satisfies a Hoelder condition of order $\alpha$ (i.e. functions of $\mathcal{C}^{k}_{\alpha}$).

\section{FIR Approximation}
\label{sec:FIR_Approximation}

Corollary~\ref{cor:WienerFilter_smooth} shows that if $\Phi$ and $\Psi$ are sufficiently smooth, the Wiener filter $H$ belongs to $\AD$ and can therefore be approximated uniformly by a polynomial. The minimal error (measured in the stability norm) which can be achieved by approximating $H \in \AD$ by an polynomial of degree $N$ ($\mathcal{P}_{N}$ denotes the set of all such polynomials) is denoted by 
\begin{eqnarray*}
	& \epsilon_{N}(H) = \inf_{P\in\mathcal{P}_{N}} \left\| H-P \right\|_{\infty}.
\end{eqnarray*}
$\epsilon_{N}(H)$ is called the \emph{best approximation} of $H$ of order $N$, and due to a theorem of Weierstrass, $\epsilon_{N}(H)$ tends to zero as $N\rightarrow\infty$ for all $H\in\AD$. Moreover, for functions which satisfy additionally a Hoelder condition of a certain order $\alpha$, the order with that $\epsilon_{N}$ goes to zero is known \cite{Zygmund}:

\begin{lemma}[Jackson,1912]
\label{lem:Jackson}
Let $H \in \A_{\alpha}(\D)$ then $\epsilon_{N}(H) \leq C N^{-\alpha}$ with a constant $C$ independent on $N$.
And if to a given $H$ there exists a polynomial approximation of degree $N$ such that $\epsilon_{N}(H) \leq C N^{-\alpha}$, then $H \in \A_{\alpha}(\D)$.
\end{lemma}

Thus, as smoother $H$ (i.e. as larger $\alpha$) as faster decreases the approximation error as $N$ is increased.

The question is now: Which polynomial of degree $N$ achieves the best approximation $\epsilon_{N}(H)$?
Or at least, which polynomials show the approximation behavior of Lemma~\ref{lem:Jackson}?
To answer this question, we write $H \in \AD$ as $H(z) = \sum^{\infty}_{k=0}H_{k}z^{k}$ and consider the so called \emph{Fejér mean} of $H$ (see e.g. \cite[chapter~3]{Zygmund})
\begin{eqnarray*}
	& F_{N}(z;H) = \sum^{N}_{k=0} \left(1 - \frac{k}{N+1}\right)H_{k}z^{k}\ .
\end{eqnarray*}
Therewith a class of approximation polynomials can be defined, for which the error decreases as in Lemma~\ref{lem:Jackson}: 
\begin{equation}
\label{equ:ValleePoussinMean}
V_{N}(z;H) = 2 F_{2N-1}(z;H) - F_{N-1}(z;H)\ .
\end{equation}
They are called the \emph{de la Vallée-Poussin means} of $H(z)$ and the following result is well known. \cite[chapter~3]{Zygmund}
\begin{lemma}
Let $V_{N}$ be the de~la~Vallée-Poussin mean of $H$, then
$\left\|H - V_{N}\right\|_{\infty} \leq 4\ \epsilon_{N}(H)$.
\end{lemma}

For $H \in \A_{\alpha}(\D)$ follows (together with Lemma~\ref{lem:Jackson}) for the approximation error that $\left\|H - V_{N}\right\|_{\infty} \leq 4 C N^{-\alpha}$. Moreover the approximation error of the de la Vallée-Poussin mean $V_{N}(z,H)$ is upper bounded by the best approximation $\epsilon_{N}(H)$, although it should be noted that $V_{N}(z;H)$ is a polynomial of order $2N-1$.
The consequences for our three filters are summarized in the following two theorems:
\begin{theorem}
\label{theo:ApproxGeschw}
Let $\Phi,\Psi \in \mathcal{C}_{\alpha}[-\pi,\pi)$ and let $H$ be the transfer function of the whitening-, of the estimation- \eqref{equ:ElementarFilter}, or of the Wiener filter \eqref{equ:WienerSolution}.
Then, there exists an FIR approximation $H_{N}$ of degree $N$ such that 
	$\left\|H - H_{N}\right\|_{\infty} \leq C N^{-\alpha}$
with a constant $C$ independent of $N$. Moreover, this approximation velocity is achieved by using de~la~Vallée-Poussin means $V_{N}(z,H)$ of $H$ as approximation polynomials $H_{N}$.
\end{theorem}

With the previous discussion, this theorem is already proofed. For the whitening filter $H_{W}$ also the reverse is true:
\begin{theorem}
Let $H_{W}$ be the whitening filter \eqref{equ:whiteningFilter} corresponding to a given spectral density function $\Phi$ and let $H_{N}$ be an FIR approximation of degree $N$ such that
\begin{equation*}
	\left\|H_{W} - H_{N}\right\|_{\infty} \leq C N^{-\alpha}
\end{equation*}
then $\Phi\in \mathcal{C}_{\alpha}[-\pi,\pi)$.
\end{theorem}

Altogether this shows that the smoothness of $\Phi$ and $\Psi$ is directly related to the rate with that the approximation error approaches zero. As smoother the densities are, as shorter the approximative FIR filter can be.

However, if the spectral densities are not Hoelder continuous, the corresponding filters can not be approximated by an FIR filter. Moreover, it can be shown by using Theorem~\ref{DiskAlgebra} and Corollary~\ref{Corollary2} that there exist spectral densities $\Phi\in\mathcal{C}_{+}\left[-\pi,\pi\right)$ such that the best approximation of its outer function $F_{+}$, using an approximation function $g\in\AD$ from the disk algebra is
\begin{eqnarray*}
	\epsilon\left(F_{+}\right) = \inf_{g\in\AD}\left\|F_{+}-g\right\|_{\infty}
	                           = \left\|F_{+}\right\|_{\infty}
\end{eqnarray*}
and that this minimal approximation error is achieved by $g\equiv0$.
Of course this means that $F_{+}$ can actually not be approximated by a function from $\AD$ and therefore it can not be approximated by an FIR filter.

Finally, we mention that the methods of this paper can also be used to study the continuous dependency of the Wiener filter $H$ on the given data $\Phi$ and $\Psi$. First results show that every $\Phi \in \mathcal{C}_{+}[-\pi,\pi)$ is a \emph{discontinuity point of the spectral factorization}. This means that for every $\widetilde{\Phi}$ with $\|\Phi-\widetilde{\Phi}\|_{\infty}<\epsilon$, the relative error in the spectral factors $\frac{1}{\epsilon}\|\Phi_{+}-\widetilde{\Phi}_{+}\|_{\infty}$ may becomes arbitrarily large. Similar effects appear also in FIR filter design. There, the error grows proportional to $\log N\cdot\epsilon$ where $N$ is the FIR filter degree. A detailed investigations will be published elsewhere.

An extension of the investigations of this paper to multiple-input multiple-output systems will be given in \cite{Boche_Pohl_PIMRC05,Boche_Pohl_WPMC05}.

\section{Conclusions}

For a Wiener filter, it is sufficient that the given spectral density functions $\Phi$ and $\Psi$ satisfy a Hoelder condition, in order that the Wiener filter is always stable and can be uniformly approximated by an FIR filter. Moreover the smoothness of $\Phi$ and $\Psi$ characterize how fast the approximation error decreases as the degree of the approximation filter is increased. If $\Phi$ and $\Psi$ are only continuous, the estimation filter may become unstable and it may not be possible to approximate the whitening filter by an FIR filter. It can be shown that this holds also for absolute continuous spectra.


\end{document}